\documentstyle[amsmath,amssymb,aps,multicol,epsf,prl]{revtex}

\draft

\newcommand{\figureone}{
\begin{figure}
\epsfxsize=0.9 \linewidth
\epsfbox{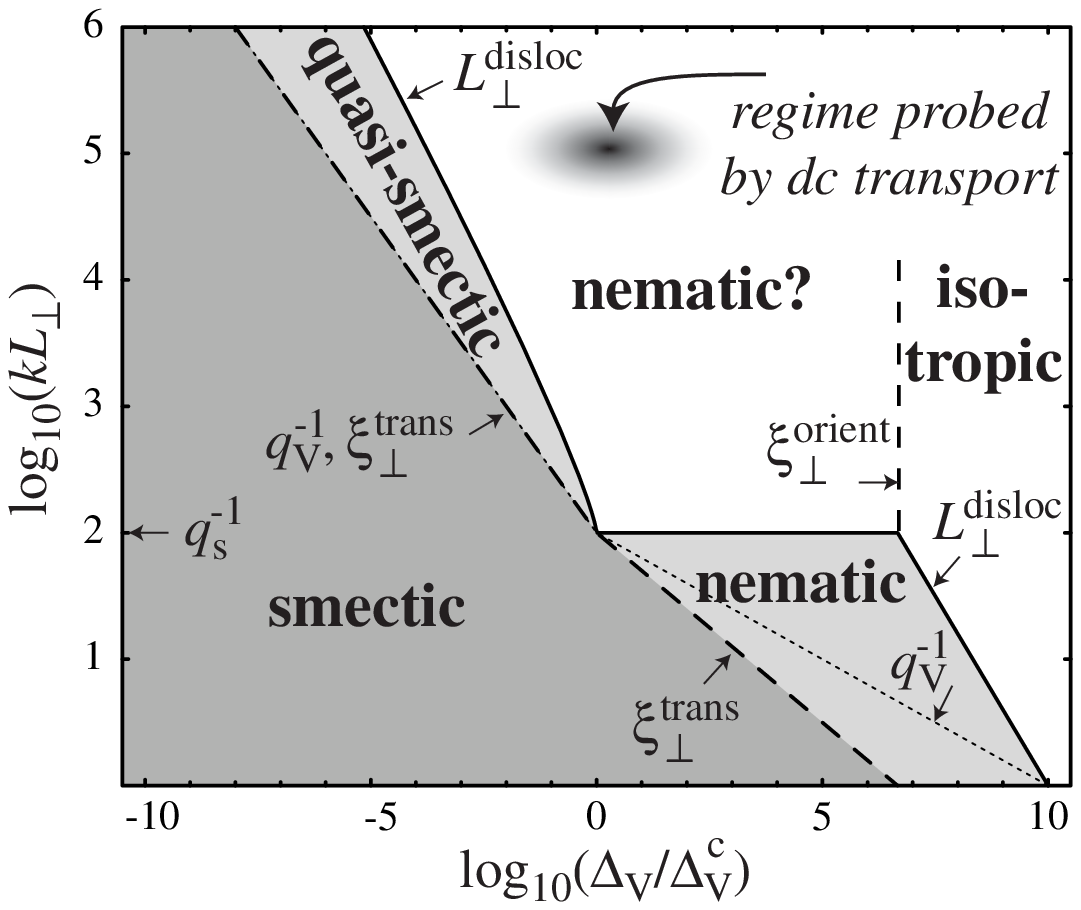}
\narrowtext
\caption{
  Illustration of regimes in dependence of disorder strength $\Dv$ and
  transverse length scale $\Lpe$ for $\qs \approx 0.01 k$ as is
  typical for experiments. In the smectic regime $\Lpe \ll \xitpe$
  translational and orientational order are well pronounced, $f \ll
  k^{-2}$ and $g \ll 1$. In the weak-disorder case, $\Dv \ll \Dv^\cc$,
  this is followed by a quasi-smectic regime with quasi-long-range
  translational order.  In the strong-disorder case, $\Dv \gg
  \Dv^\cc$, a nematic regime follows, where translational order is
  effectively short ranged.  The strict validity of our analysis based
  on elastic theory ends at $\Lpe=\Ldis$, where free dislocations
  appear.  Nevertheless, since in the absence of dislocations
  orientational order would be long-ranged (i.e. $\xiope = \infty$;
  except for extremely strong disorder), nematic order possibly
  persists even for $\Lpe \gg \Ldis$. dc transport measurements on
  macroscopic samples probe the regime in the indicated area.}
\label{fig}
\end{figure}
}

\newcommand{\qs}{q_{\rm s}}
\newcommand{\qv}{q_V}

\newcommand{\cc}{{\rm c}}

\newcommand{\Dh}{\Delta_h}
\newcommand{\Dv}{\Delta_V}

\newcommand{\xitpe}{\xi^{\rm trans}_\perp}
\newcommand{\xiope}{\xi^{\rm orient}_\perp}

\newcommand{\be}{{\bf e}}

\newcommand{\bq}{{\bf q}}
\newcommand{\br}{{\bf r}}

\newcommand{\rpa}{r_\parallel}
\newcommand{\rpe}{r_\perp}

\newcommand{\qpe}{q_\perp}

\newcommand{\Epin}{E_{\rm pin}}
\newcommand{\Eself}{E_{\rm self}}

\newcommand{\Bpa}{B_\parallel}
\newcommand{\Bpe}{B_\perp}

\newcommand{\Lpe}{L_\perp}

\newcommand{\Ldis}{L^{\rm disloc}_\perp}
\newcommand{\ndis}{n_{\rm disloc}}

\begin{document}

\title{Elastic theory of quantum Hall smectics: effects of disorder}

\author{Stefan Scheidl and Felix von Oppen}

\address{Institut f\"ur Theoretische Physik, Universit\"at zu
  K\"oln, Z\"ulpicher Str. 77, D-50937 K\"oln, Germany}

\date{\today}
\maketitle

\begin{abstract}
  We study the effect of disorder on quantum Hall smectics within the
  framework of an elastic theory. Based on a renormalization group
  calculation, we derive detailed results for the degrees of
  translational and orientational order of the stripe pattern at zero
  temperature and carefully map out the disorder and length-scale
  regimes in which the system effectively exhibits smectic, nematic,
  or isotropic behavior.  We show that disorder always leads to a
  finite density of free dislocations and estimate the scale on which
  they begin to appear.
\end{abstract}


\begin{multicols}{2} \narrowtext

Recent experiments \cite{Lilly+99:a} discovered a pronounced
anisotropy in the resistivity of the quantum Hall system near half
filling of higher Landau levels (Landau level filling factor
$\nu\ge3$), suggesting that the ground state at these filling
factors breaks orientational symmetry. It is believed that this is a
manifestation of the formation of striped quantum Hall states whose
existence had previously been predicted based on Hartree-Fock (HF)
calculations \cite{HF1} (for a recent review, see \cite{vOp99}). The
existence of striped states in higher Landau levels has also been
supported by exact numerical diagonalization studies \cite{ReHa99}
and by the experimental confirmation \cite{Lilly+99:a} of some
explicit predictions \cite{MF00,vOp00} for their transport
properties.

The HF calculations predict that the ground state near half filling of
higher Landau levels is a unidirectional charge density wave (CDW)
whose period $a$ is of the order of the cyclotron radius \cite{foot1}.
In addition to orientational order, this state also breaks
translational symmetry in one direction. For this reason, it is often
referred to as a quantum Hall smectic state.  Going beyond mean-field
level by including quantum and thermal fluctuations, one expects a
variety of phases with different degrees of translational and
orientational order\cite{FK99,MF00}.

Here we study the effect of disorder on quantum Hall smectics at zero
temperature within the framework of an elastic theory. As a function
of disorder strength, we map out in detail various regimes in which
the system effectively exhibits smectic (with both translational and
orientational order), nematic (with orientational order only), or
isotropic behavior.  A concise summary of our results is provided in
Fig.\ 1.

At low energies, the relevant gapless excitations are fluctuations of
the displacement field $u(\br,\tau)$ associated with translational
symmetry breaking \cite{MF00,vOp99} and defined as the phase of the
charge-density modulation $\rho(\br,\tau)=\rho_0\cos
[k(r_\parallel+u(\br,\tau))]$.  The position vector $\br=(\rpa,\rpe)$
is represented by its components parallel and perpendicular to the CDW
wave vector ${\bf k}$ with $k=2\pi/a$.  We note that as opposed to the
stripe position described by $u$, the stripe width is a massive mode
that can be neglected.

\figureone

The chiral nature of quantum Hall systems leads to a dynamic behavior
of quantum Hall smectics which differs in important ways from that of
ordinary smectic liquid crystals.  The elastic action for quantum Hall
smectics takes the form \cite{MF00} ($\beta$ denotes the inverse
temperature)
\begin{equation}
  S_{\rm el}[u]={1\over \beta}\sum_\omega \sum_\bq M
  {\omega^2\over q_\perp^2}
  |u(\bq,\omega)|^2+\int_0^{\beta\hbar} d\tau\, E(\tau),
\end{equation}
where $M$ is a dynamic coefficient and $E(\tau)$ denotes the elastic
free energy. The inverse powers of the transverse wave vector $\qpe$
in the dynamic term are {\it specific} to quantum Hall smectics. Most
importantly, they imply that smectic order is stable against (weak)
quantum fluctuations at zero temperature in the quantum Hall system,
even in the absence of an alignment force orienting the stripes along
a preferred direction \cite{MF00}.

The elastic free energy is of the smectic form \cite{MF00}
\begin{eqnarray}
  E(\tau)&=&{1\over2}\int_\br \bigg[\Bpa\left(\nabla_\parallel u+{1\over2}
    (\nabla_\perp u)^2+ \frac C2\right)^2
  \nonumber\\
  &&\,\,\,\,\,\,\,\,\,\,\,\,\,\,\,\,\,\,\,\,\,\,\,\,\,\,\,\,\,\,
  +K(\nabla_\perp^2 u)^2+\Bpe(\nabla_\perp u)^2\bigg].
  \label{free-energy}
\end{eqnarray}
The constant $C$ is determined by the requirement $\langle
\nabla_\parallel u\rangle=0$ which fixes the CDW wave length.  The
compression modulus $\Bpa$ and the bending modulus $K$ are related by
$\Bpa=4 k^2 K$ \cite{GP81}.  For $B_\perp=0$, this free energy is
rotation invariant due to the anharmonic terms \cite{GP81}.  We have
included a crystal field with tilt modulus $\Bpe$ which accounts for
the presence of an alignment force orienting the stripe pattern along
a preferred direction.  Experiments suggest that this alignment force
is quite weak, $\Bpe \ll \Bpa$.  The presence of the tilt modulus
leads to an important (transverse) length scale in the problem,
defined by $\qs=(\Bpe/K)^{1/2}$.  For $\qpe \gg \qs$, the bending
energy of the stripes dominates over the tilt energy while for $\qpe
\ll \qs$, the tilt energy dominates.

Various kinds of disorder may affect the quantum Hall stripes.  A
disorder potential $U(\br)$ is included via an impurity action
$S_{{\rm imp},V}=\int_{\br,\tau} U(\br) \rho(\br,\tau)$. It can be
rewritten as $S_{{\rm imp},V}=(\rho_0/2) \int_{\br,\tau}
[V(\br)e^{iku(\br,\tau)} + \textrm{c.c.}]$, where $V$ is effectively
locally correlated, $\langle V(\br)V^*(\br') \rangle\simeq
W\delta(\br-\br')$, even for long-range-correlated impurity potentials
$U(\br)$ because of the oscillatory nature of the charge density
$\rho(\br,\tau)$.  The effective disorder strength $W$ is given by $W
\simeq \langle U(-{\bf k})U({\bf k})\rangle$.  For quantitative
estimates, we employ the impurity potential created by the ionized
impurities (of areal density $n_{\rm imp}$) in the spacer layer which
dominates the transport properties of two-dimensional electron systems
(2DES). For uncorrelated impurity positions, the disorder potential
experienced by the electrons in the 2DES has $W=e^2 n_{\rm
  imp}v_d^2(k)$.  Here, $v_d(q)=2\pi e^{-qd}/\epsilon(q) q$ is the
screened Coulomb interaction, $\epsilon(q)$ the dielectric function,
and $d$ the distance of the impurity layer from the 2DES.

While the bare action contains only the point disorder $U(\br)$, the
action of the renormalization group (RG) generates a second type of
disorder.  This effective disorder corresponds to a random tilting
force acting on the stripe pattern and is described by the additional
term $S_{{\rm imp},h}=\int_{\br,\tau}h(\br)\nabla_\perp u(\br,\tau)$
in the action with a local correlator $\langle h(\br)h(\br')\rangle =
\Delta_h \delta(\br-\br')$.

We proceed in a standard fashion by averaging the free energy over the
disorder potential using the replica trick which leads to the
impurity-averaged replicated partition function
\begin{eqnarray}
  \langle Z^n\rangle&=&\int \prod_\alpha [du_\alpha] \exp 
  \bigg\{-{1\over\hbar}
    \sum_\alpha S_{\rm el}[u_\alpha]
    \nonumber\\
   && +{1\over2\hbar^2}\int_{\br,\tau,\tau'}\sum_{\alpha,\beta}\Big[
   \Delta_h(\nabla_\perp u_\alpha(\br,\tau))(\nabla_\perp
   u_\beta(\br,\tau'))
   \nonumber\\
   && +\Delta_V\cos[k(u_\alpha(\br,\tau)-u_\beta(\br,\tau'))]\Big]\bigg\}.
\label{av-partition}
\end{eqnarray}
Here, $\Delta_V=(1/2)\rho_0^2W$ measures the strength of the point 
disorder while $\Delta_h$ denotes the strength of the tilt disorder.

Since translational order is stable against quantum fluctuations at
zero temperature\cite{MF00}, one might attempt to treat the disorder
potential perturbatively by expanding the action in
Eq.~(\ref{av-partition}) to quadratic order in $u$, setting
$\Delta_h=0$, and neglecting dislocations.  In the resulting quadratic
theory it is straight forward to calculate the correlation functions
\begin{mathletters}
\begin{eqnarray}
  f(\br)&=&\langle [u(\br')-u(\br'+\br)]^2 \rangle ,
  \\
  g(\br)&=&\langle [\nabla_\perp' u(\br')-\nabla_\perp' u(\br'+\br)]^2 
  \rangle ,
\end{eqnarray}
\end{mathletters}
which describe the degrees of translational and orientational order,
respectively.  Neglecting quantum fluctuations which are additive in
perturbation theory, one finds at zero temperature $f(\br_\parallel) =
{1\over 2} \int_{q_\perp} ({k^2\Dv / \Bpa^2 Q_\parallel^3})
[1-(1+Q_\parallel r_\parallel ) e^{-Q_\parallel r_\parallel}]$ and
$f(\br_\perp) = {1\over 2}\int_{q_\perp} ({k^2 \Dv / \Bpa^2
  Q_\parallel^3}) [1-\cos(q_\perp r_\perp)]$.  Here $\br_\parallel$
($\br_\perp$) are vectors in the parallel (perpendicular) direction
and $Q_\parallel(\qpe) =[(\Bpe q_\perp^2 + Kq_\perp^4)/\Bpa]^{1/2}$.
The analogous expressions for $g(\br)$ differ from those for $f(\br)$
merely by an additional factor $q_\perp^2$ in the integrand.

One readily finds that $f(\br)$ is infrared divergent for any nonzero
$\br$, irrespectively of whether the tilt modulus $B_\perp$ vanishes
or not.  Because of the multistability of the disorder potential,
perturbation theory breaks down on scales where $f \gtrsim a^2$
\cite{FLRL}, i.e.  {\em on all finite scales}.  Below, we perform a RG
calculation which properly accounts for the multistability by a
significant downward renormalization of the disorder strength on large
scales.

The perturbative expression can nevertheless be used to estimate the
characteristic transverse scale $\qv$, below which the renormalization
sets in. To do so, we demand $f(\qv^{-1}\be_\perp) \sim a^2$ with the
provision that the integral over $\qpe$ in the perturbative expression
for $f$ be self-consistently restricted to $\qpe \geq \qv$. The
explicit expression for $\qv$ depends on the disorder strength.  In
the {\it weak-disorder regime} $\qv\ll\qs$, one finds
$\qv=[\Dv/\Dv^\cc]^{1/2} \qs$, while in the {\it strong-disorder
  regime} $\qv \gg \qs$ one obtains $\qv=[\Dv/\Dv^\cc]^{1/5} \qs$ ,
The disorder strength separating these two limits is given by
$\Delta_V^\cc= 4 \pi \Bpa^{1/2} \Bpe^{5/2}/k^4 K$.
 
We now turn to the RG calculation, restricting ourselves to zero
temperature where the smectic state is stable in the pure system. As
the approach neglects topological defects such as dislocations, the
results are valid only up to the scale $\Ldis$ where dislocations
start to appear. This length will be estimated below.  Our analysis
builds on previous RG studies for classical smectics \cite{GP81,RT99},
adapting them to two dimensions where disorder is known to play a
peculiar role \cite{VF84}.  We employ a conventional
coarse-graining procedure, where all displacement modes in the range
$\Lambda<|q_\perp|<k$ with $\Lambda=ke^{-l}$ are integrated out.  (A
length rescaling is omitted.)  Treating both the point disorder and
the anharmonic terms in the elastic action to one-loop order leads to
the flow equations
\begin{mathletters}
\label{anharmonic-flow}
\begin{eqnarray}
  {d\Delta_V\over dl}&=& -a_1 \Dv ,
  \\
  {d\Delta_h\over dl}&=&  a_2 \Dv + a_3 \Dh ,
  \\
  {d\Bpa\over dl}&=& - 3 a_3 \Bpa,
  \\
  {dK\over dl}&=&  2 a_3 K,
\end{eqnarray}
\end{mathletters}
where $a_1(\Lambda) = k^4 \Lambda\Delta_V/4\pi
\Bpa^2Q_\parallel^3(\Lambda)$ and $a_2(\Lambda)= c (k^2/\Lambda^2)
a_1(\Lambda)$ with $c$ a constant of order unity.  Moreover,
$a_3(\Lambda)=\Bpa^{1/2}\Delta_h/16\pi K^{5/2}\Lambda^3$ for
$\Lambda\gg\qs$ and $a_3=0$ for $\Lambda\ll\qs$. The couplings $a_1$
and $a_2$ arise from the point disorder while $a_3$ is due to the
anharmonic terms in the elastic action.  We note that the tilt modulus
$\Bpe$ remains unrenormalized.  The renormalization of $\Bpa$ and $K$
is due to the combined effect of disorder and anharmonic elasticity.
The characteristic disorder scale $\qv$ can be recovered from the flow
equations by the condition $a_1(\qv) = 1$ which defines the effective
onset of the renormalization of $\Delta_V$.  We first solve the flow
equations ignoring the anharmonic terms in the elastic action (i.e.
$a_3=0$).  As discussed below, these terms should lead only to minor
modifications.  Solving the flow equation we find that the effective
disorder strength $\Delta_V \sim \Lambda^2$ flows to zero for $\Lambda
\ll \qv$, while the tilt disorder strength $\Delta_h$ increases
logarithmically.  Due to this decay of $\Dv$ the coupling $a_1$
remains finite for vanishing $\Lambda$, which indicates that the
renormalization of disorder is captured consistently on one-loop
level.

The detailed results depend on the disorder strength. In the {\it
  weak-disorder regime} $\qv\ll\qs$ or $\Dv \ll \Dv^\cc$, we find for
the translational order in the soft (perpendicular) direction
\begin{equation}
  f(\br_\perp)\sim {1\over k^2}
  \left\{
    \begin{array}{ll}
      (\qv r_\perp)^2 \ln(\qs/\qv) ,
      &
      r_\perp \ll \qs^{-1} ,
      \cr
      (\qv r_\perp)^2 \ln(1/\qv r_\perp) ,
      & 
      \qs^{-1}\ll r_\perp \ll \qv^{-1} ,
      \cr
      \ln^2(\qv r_\perp),
      &  
      r_\perp \gg \qv^{-1}.
    \end{array}\right.  
\label{f.clean}
\end{equation} 
As opposed to the perturbative result, $f$ is now finite for any
finite $r_\perp$ and we can identify a correlation length $\xi^{\rm
  trans}_\perp$ for the translational order by $f(\xi^{\rm
  trans}_\perp)=a^2$ and find $\xi^{\rm trans}_\perp \sim \qv^{-1}$.
The slow $\ln^2$ increase of $f$ in the regime $r_\perp\gg\xi^{\rm
  trans}_\perp$, typical of two-dimensional disordered systems
\cite{VF84}, implies that the system exhibits ``quasi''-long-range
translational order, with power-law Bragg peaks with logarithmically
varying powers.  The orientational correlation function $g$ remains
finite, $g(\br_\perp)\sim (\qv/ k)^2\ln(\qs/\qv) \ll 1$, for
arbitrarily large lengths $r_\perp \gg \qv^{-1}$.  Hence the
correlation length $\xi^{\rm orient}_\perp$ for orientational order,
defined by $g(\xi^{\rm orient}_\perp)=1$, is infinitely large.

For sufficiently strong disorder, we enter the {\it strong-disorder
  regime} $\qv\gg\qs$ or $\Dv \gg \Dv^\cc$ where
\begin{eqnarray}
  f(\br_\perp) \sim {1\over k^2} 
  \left\{
    \begin{array}{ll}
      (\qv/\qs)^3(\qs r_\perp)^2 ,
      & 
      r_\perp \ll \qs^{-1} ,
      \cr
      (\qv/\qs)^3\ln(\qs r_\perp),
      & 
      \qs^{-1} \ll r_\perp \ll q_h^{-1},
      \cr
      \ln^2(\qs\rpe),
      &
      \rpe\gg q_h^{-1}.
    \end{array}\right.  
\label{f.dirty.harm}
\end{eqnarray}
In this case, the $\ln^2$ regime typically sets in only at extremely
large length scales $\rpe\gg q_h^{-1}$ with
$q_h=\qs\exp[-(\qv/\qs)^3]$ so that we can restrict attention to the
$\ln$ regime. At first sight, this seems to suggest the presence of
quasi-long-range translational order in this case. While this is
formally true, the corresponding singularities in the structure factor
are extremely weak due to the large prefactor $(\qv/\qs)^3$ and for
practical purposes, there is only short-range translational order with
correlation length $\xi^{\rm trans}_\perp=(\qs/\qv)^{1/2}\qv^{-1}$.
As expected, the degree of translational order is significantly
reduced compared to the weak-disorder regime.  The degree of
orientational order is also reduced, since $g(\br_\perp)\sim (\qv/
k)^2(\qv/\qs)$ for $r_\perp\gg \qs^{-1}$.  Nevertheless, $\xiope
=\infty$ as long as $\qv^3 \ll k^2 \qs$.  Otherwise, for extremely
strong disorder, $\xiope \sim k^2 / \qv^3$.

Results for the correlation functions in the parallel direction in the
asymptotic regimes can be easily obtained from those in the
perpendicular direction by the replacement $r_\perp\to
(\qs/k)r_\parallel$. The small factor $(\qs/k)\ll1$ reflects the fact
that the system is more rigid in the parallel than in the
perpendicular direction.  It is interesting to note that in the
absence of an alignment force ($\Bpe=0$) orienting the stripes along a
preferred direction, disorder destroys translational order and leads
to quasi-long-range orientational order, even when neglecting
dislocations.

So far the density modulation has been treated assuming the absence of
any topological defects. We now show that disorder always induces a
finite density of dislocations and estimate the transverse scale
$\Ldis$ on which dislocations become relevant and which limits the
range of validity of the above results.  The occurrence of
dislocations depends on a competition between the elastic energy cost
$\Eself$ to create a dislocation and the energy gain $\Epin$ in the
disorder potential. The energy gain $\Epin$ originates from the
reduced rigidity of the charge density profile in the presence of
dislocations.  The scale $\Ldis$ is identified as the size of a finite
system where the free energy $\Eself+\Epin$ of a single dislocation
becomes negative.  For a strictly smectic system ($\Bpe=0$), the
elastic energy cost $\Eself$ of a dislocation is known to be finite.
A finite tilt modulus $\Bpe$ leads to a logarithmic increase in the
elastic energy with system size $\Lpe$ for $\Lpe \gg \qs^{-1}$.  The
energy of the dislocation in the disorder potential is a random
variable that depends on the location of the dislocation.  The pinning
energy $\Epin$ is given by the {\it maximal} energy that the
dislocation can gain from the disorder potential. Assuming a Gaussian
distribution for the disorder energy, this energy can be computed
using the statistics of extreme values.  In the {\it weak-disorder
  regime}, we eventually find that the creation of dislocations
becomes favorable on transverse scales
\begin{eqnarray}
  \Ldis \sim \qv^{-1} \exp[c \ln^{2/3}(\qs/\qv)]
  \label{Ldis.clean}
\end{eqnarray}
(with $c$ a constant of order unity) corresponding to an areal
dislocation density of $\ndis \sim (\Ldis)^{-2} (\Bpe/\Bpa)^{1/2}$.
In the {\it strong-disorder limit}, we find
\begin{eqnarray}
  \Ldis \sim \qs^{-1} \min [1, k^2\qs /\qv^3]
\end{eqnarray}
and a dislocation density of $\ndis \sim(\Ldis)^{-3} (K/\Bpa)^{1/2}$.
Thus, within a finite range of disorder strengths, $\Dv^\cc \ll \Dv
\ll (k/\qs)^{10/3} \Dv^\cc$ or $\qs^3 \ll \qv^3 \ll k^2 \qs$, $\Ldis$
saturates at $\qs^{-1}$, before it shrinks further, coinciding with
$\xiope$.

The present analysis of regimes neglects the influence of the
anharmonic elastic terms, although they are included at one-loop level
in the flow equations (\ref{anharmonic-flow}). On scales $\qpe \ll
\qv$ these terms lead to an effective reduction of $\Bpa$ and an
increase of $K$, making the system appear more isotropic.  The
integration of the full flow equations \cite{unpub} modifies the
correlation functions {\em only slightly} in the strong-disorder
limit.  In addition, $\qs$ is reduced due to the renormalization of
$K$.

Our results therefore lead to the following overall picture, as
summarized in Fig.\ 1. In the clean limit we have a sequence of
transverse length scales $\qs^{-1} \ll \qv^{-1} \simeq \xitpe \ll
\Ldis$ (see Fig. \ref{fig}). On scales $\Lpe \ll \qv^{-1}$, the system
looks like a {\em smectic} with pronounced translational order ($f \ll
a^2$) and no free dislocations.  On scales $\qv^{-1} \ll \Lpe \ll
\Ldis$, there are still no free dislocations. However, $f \gg a^2$ and
$f$ increases logarithmically, displaying quasi-long-range order,
whence we call this regime {\em quasi-smectic}.  For $\Lpe \gg \Ldis$
dislocations appear and destroy translational order.  Neglecting
dislocations, one finds $g \ll 1$ (i.e. $\xiope = \infty$) on all
scales in the weak-disorder regime, which indicates the possibility to
have good orientational (nematic) order even for $\Lpe \gg \Ldis$.
 
In the strong-disorder regime, the sequence of scales is $\xitpe \ll
\qv^{-1} \ll \Ldis \lesssim \qs$. Good translational order survives
only for $\Lpe \ll \xitpe$. Beyond this scale, translational order is
rapidly lost as $f$ increases as a power law with $f\gg1$.  On the
other hand, orientational order is still retained, $g \ll 1$.  Hence,
we cross over from a {\it smectic} regime for $\Lpe \ll \xitpe$ to a
{\em nematic} regime for $\Lpe \gg \xitpe$. Dislocations are present on
scales $\Lpe \gg \Ldis$. For $\Dv^\cc \ll \Dv \ll (k/\qs)^{10/3}
\Dv^\cc$, $\xiope =\infty$ in the absence of dislocations and nematic
order may survive beyond $\Ldis$ even in their presence.  For $\Dv
\gg (k/\qs)^{10/3} \Dv^\cc$, $\xiope \sim \Ldis$ and most likely
nematic order is destroyed for $\Lpe \gg \xiope$, leading to an
isotropic regime.

For typical experimental samples \cite{Lilly+99:a}, we estimate $\qs
\sim \qv \sim 0.01 k$, indicating that they are located in the
crossover region between the weak- and strong-disorder regimes (cf.
Fig.\ 1).  While we cannot firmly deduce the presence of orientational
order on the macroscopic scales probed by {\em dc} transport
measurements, there is a good chance that orientational order persists
even for $\Lpe \gtrsim \Ldis$, since $\xiope =\infty$ for $\Dv \ll
(k/\qs)^{10/3} \Dv^\cc$ in the absence of dislocations.  The regimes
identified in this paper at smaller lengths may be accessible to other
experimental probes such as surface acoustic waves.

This work was partly supported by the DFG through Schwerpunktprogramm
``Quanten-Hall-Systeme'' and Sonderforschungsbereich 341.

\end{multicols}

\begin{references}
\bibitem{foot1} In principle, the CDW is unstable with respect to
  charge ordering along the stripes at extremely low temperatures
  \cite{MF00}.  Presumably this ordering is destroyed by minute
  impurity concentrations even at zero temperature. Thus we will
  ignore this complication.
\end{references}
\end{document}